\documentclass[a4paper,11pt]{article}

\usepackage{jheppub}

\title{\boldmath Spacetime dilaton in ${\rm AdS}_3\times X$ holography}

\abstract{We study the spectrum generating operators (DDF operators) for bosonic strings and superstrings in ${\rm AdS}_3\times X$ with pure, \emph{generic} NSNS flux. The dual CFT is conjectured to be a deformed symmetric orbifold of $\mathbb{R}_{\widetilde{Q}}\times X$, where $\mathbb{R}_{\widetilde{Q}}$ denotes a linear dilaton theory with slope $\widetilde{Q}$. In this paper, we construct DDF operators corresponding to the spacetime linear dilaton field in the limit where the worldsheet theory becomes free. The DDF operator gives another evidence supporting the proposed dual CFT. Furthermore, the DDF operator is constructed purely from the $\rm AdS_3$ fields and thus, strongly implies the existence of $\mathbb{R}_{\widetilde{Q}}$ for generic $X$ that defines a consistent background.}

\author{Vit Sriprachyakul}
\affiliation{Institut f\"ur Theoretische Physik,
ETH Z\"urich,\\
Wolfgang-Pauli-Strasse 27,
8093 Z\"urich, Switzerland}
\emailAdd{vsriprachyak@phys.ethz.ch}

\usepackage{bm} 
\usepackage{xcolor}
\definecolor{green_maf}{RGB}{28, 166, 46}
\definecolor{blue_mrg}{RGB}{12, 143, 145}
\definecolor{detail}{RGB}{110,110,110}
\usepackage{amsmath} 
\usepackage{nccmath} 
\usepackage{amssymb} 
\usepackage{amsthm} 
\usepackage{mathtools} 
\usepackage[utf8]{inputenc} 
\usepackage{braket}
\usepackage{enumerate}
\usepackage[inline]{enumitem}
\usepackage{comment}
\usepackage{soul} 
\usepackage{tikz}
\usepackage{csquotes}

\usetikzlibrary{calc,decorations.markings}
\usetikzlibrary{decorations.pathmorphing}
\usetikzlibrary{shapes,backgrounds}
\usetikzlibrary{fadings}
\usetikzlibrary{cd}
\usetikzlibrary{decorations.pathreplacing,calligraphy}

\tikzset{
partial ellipse/.style args={#1:#2:#3}{
insert path={+ (#1:#3) arc (#1:#2:#3)}
}
}

\newif\ifdetails
\detailstrue

\def\be{\begin{equation}}
\def\ee{\end{equation}}

\begin{document}

\maketitle

\section{Introduction}

There has been a lot of progress in $\rm AdS_3/CFT_2$ recently and some of the most interesting ones are tensionless holography and the near-boundary limit. The common feature that makes these examples appealing is simultaneous tractability on both sides of the duality. In the tensionless limit, both the worldsheet theory and the dual CFT are free. This allows us to perform many explicit tests of this particular $\rm AdS_3/CFT_2$. We will not discuss tensionless holography in this paper but interested readers may consult a non-exhausive list \cite{Eberhardt:2018ouy,Eberhardt:2019ywk,Dei:2020zui,Fiset:2022erp,Gaberdiel:2021kkp,Knighton:2024noc,Dei:2023ivl,Eberhardt:2020bgq,Gaberdiel:2023lco} and references therein. The near-boundary limit involves focusing on a specific configuration of the worldsheet, that is, on the worldsheet that localises near the conformal boundary of $\rm AdS_3$. It turns out that doing so allows one to explore strings in $\rm AdS_3$ with pure, generic NSNS flux while still retaining similar analytic control as the tensionless limit. In fact, the action describing such near-boundary worldsheet is free \cite{Giveon:1998ns} and the worldsheet amplitudes can be computed explicitly \cite{Knighton:2024qxd,Sriprachyakul:2024gyl}. The dual CFT to the near-boundary worldsheet theory, on the other hand, is not free in general. Nevertheless, as far as the near-boundary $\rm AdS_3$ holography is concerned, it is sufficient to treat the dual CFT deformation perturbatively \cite{Eberhardt:2021vsx,Knighton:2024qxd}.

In \cite{Eberhardt:2021vsx}, it was first argued, by computing ground state 2- and 3-point amplitudes, that the dual CFT to bosonic strings in ${\rm AdS_3}\times X$ with pure, generic NSNS flux is
\begin{equation}
{\rm Sym}^N\left( \mathbb{R}_{\widetilde{Q}}\times X \right)+\text{a twist-2 deformation}\,,
\label{eq:bosonic string duality}
\end{equation}
where we have denoted the linear dilaton theory with dilaton slope $\widetilde{Q}$ by $\mathbb{R}_{\widetilde{Q}}$. It was then argued in \cite{Knighton:2023mhq} and later completed in \cite{Knighton:2024qxd} that the duality holds for any $n$-point functions of ground states.\footnote{In \cite{Knighton:2024qxd}, the duality was shown for scattering amplitudes of states in the $\mathfrak{sl}(2,\mathbb{R})_k$ continuous representations. The states in the $\mathfrak{sl}(2,\mathbb{R})_k$ discrete representations were argued to be dual to the bound states of the spacetime CFT, see Section 4 of \cite{Eberhardt:2021vsx}.} One of the ingredients behind the success of \cite{Knighton:2024qxd} was the near-boundary approximation which allows the authors to analytically compute generic $n$-point amplitudes and verify the duality beyond $3-$point functions. Another angle of the duality was explored in \cite{Hikida:2023jyc} where the authors construct the worldsheet theory starting from the dual CFT \eqref{eq:bosonic string duality}.\footnote{The worldsheet theory constructed in \cite{Hikida:2023jyc} has one more screening operator than the worldsheet theory in \cite{Knighton:2023mhq,Knighton:2024qxd}. This technically boils down to different mechanisms for generating the twist-2 perturbation in the dual CFT.} 

Subsequently, the near-boundary consideration was generalised to superstrings in \cite{Sriprachyakul:2024gyl} and by computing 3- and 4-point ground state amplitudes. It was then argued that the dual CFT to superstrings in ${\rm AdS_3}\times X$ with pure, \emph{generic} NSNS flux is
\begin{equation}
{\rm Sym}^N\left( \mathbb{R}^{(1)}_{\widetilde{Q}}\times X \right)+\text{a twist-2 deformation}\,,
\label{eq:supersymmetric string duality}
\end{equation}
where the superscript $(1)$ emphasises that the linear dilaton theory $\mathbb{R}^{(1)}_{\widetilde{Q}}$ is now an $\mathcal{N}=1$ linear dilaton theory. The proposed dual CFT naturally reproduces existing proposals in the literature \cite{Balthazar:2021xeh,Martinec:2021vpk}.
Despite the success of the near-boundary approximation, it is still unclear how the holographic dictionary would work out for descendant states, in particular, the dilaton descendants. In contrast, the descendant states in the $X$ CFT is better understood. If the compact manifold $X$ is described on the worldsheet by a WZW model with currents $K^a$, then the spacetime CFT also possesses this symmetry and this can be realised by worldsheet DDF operators which satisfy the same Kac-Moody algebra, see eqs.(2.24) and (3.4) of \cite{Giveon:1998ns}. This then gives a precise identification between descendants of $X$ and string physical states. However, to the best of our knowledge, no worldsheet DDF operator was constructed which accounts for the spacetime dilaton excitations. Being able to do this would make the holographic dictionary more complete and this is the main task of this paper. 

The DDF operators will be constructed in Section \ref{section:DDF construction}, where interested readers are invited to skip to. More precisely, the operators are given by \eqref{eq:bosonic DDF for phi} for bosonic strings and \eqref{eq:Phi DDF for superstrings, simplified} for superstrings. The expressions involve only the $\rm AdS_3$ fields and this then signifies that the dilaton is generic in ${\rm AdS_3}\times X$ holography, at least, if one stays in the pure NSNS flux background.

The paper is organised as follows. In Section \ref{section:review ads3 strings}, we briefly review bosonic strings and superstrings in ${\rm AdS_3}\times X$ with pure NSNS flux in the limit that the worldsheet theory is described by free fields. This is the so-called near-boundary limit. In Section \ref{section:DDF construction}, we construct DDF operators associated to the spacetime dilaton field, both in the context of bosonic strings and superstrings. We finish our exposition in Section \ref{seciton:discussion} with a discussion and we collect technical computations in Appendix \ref{appendix:assorted calculations}.

\section{\boldmath Strings in \texorpdfstring{${\rm AdS}_3\times X$}{AdS3 times X}}\label{section:review ads3 strings}
In this section, we briefly discuss string theory on ${\rm AdS}_3\times X$, both in the context of bosonic string and superstring theories. We will not try to give a complete introduction to the topic but interested readers may consult a non-exhausive list \cite{Giveon:1998ns,Knighton:2023mhq,Knighton:2024qxd,Sriprachyakul:2024gyl,Eberhardt:2021vsx,Kutasov:1999xu,Hikida:2023jyc,Dei:2021xgh,Dei:2021yom,Dei:2022pkr} and references therein. We will only focus on the left-moving part of strings and we will be implicit about the compact manifold $X$. Indeed, the right-moving part can be analysed similarly and the compact manifold $X$ is described by an independent, decoupled CFT (satisfying string criticality) which will not play any role in the subsequent DDF discussion in Section \ref{section:DDF construction}.

\subsection{Bosonic strings}

Bosonic strings in ${\rm AdS}_3$ can be described in terms of the Wakimoto fields $\Phi,\beta,\gamma$ using the first order action \footnote{It was shown in \cite{Knighton:2024qxd} that in order for $\gamma$ to have poles as a covering map should, one has to include a certain screening operator into consideration. However, the inclusion of this operator does not change the free field OPEs, when treated perturbatively as done in \cite{Knighton:2024qxd}, and therefore does not change our subsequent discussion. Thus, we will ignore the presence of this term throughout this paper.}
\begin{equation}
\begin{aligned}
S_{\rm AdS_3}=\frac{1}{2\pi}\int d^2z\left( \frac{1}{2}\partial\Phi\bar\partial\Phi-\frac{QR\Phi}{4}+\beta\bar\partial\gamma+\bar\beta\partial\bar\gamma-\nu\beta\bar\beta e^{-Q\Phi} \right)\,,
\label{eq:ads full action}
\end{aligned}
\end{equation}
where $Q\in\mathbb{R}^+$. The variable $\Phi$ parametrises the radial direction of $\rm AdS_3$ and the conformal boundary is parametrised by $\gamma$ (and its right-moving analogue $\bar\gamma$) and is located at $\Phi=+\infty$. Hence, if the worldsheet stays close to the conformal boundary, then the worldsheet sigma model becomes free, that is, the term $\beta\bar\beta e^{-Q\Phi}$ is negligible. This is the subsector of the worldsheet theory we will be working with throughout our paper. It was also shown in \cite{Knighton:2024qxd} that this subsector captures the perturbative aspects of the dual CFT. 

For later convenience, we note that the OPEs can be derived from the near-boundary action which is now free and reads
\begin{equation}
\begin{aligned}
S_{\rm AdS_3, \Phi\to+\infty}=\frac{1}{2\pi}\int d^2z\left( \frac{1}{2}\partial\Phi\bar\partial\Phi-\frac{QR\Phi}{4}+\beta\bar\partial\gamma+\bar\beta\partial\bar\gamma \right)\,,
\label{eq:ads free action}
\end{aligned}
\end{equation}
and the OPEs read
\begin{equation}
\Phi(z)\Phi(w)\sim -\ln|z-w|^2,\quad \beta(z)\gamma(w)\sim-\frac{1}{z-w},\quad \bar\beta(z)\bar\gamma(w)\sim-\frac{1}{\bar z-\bar w}\,.
\label{eq: free field OPEs}
\end{equation}
Furthermore, the (left-moving) stress tensor is given by
\begin{equation}
T_{\rm AdS_3}=-\frac{1}{2}\partial\Phi\partial\Phi-\frac{Q}{2}\partial^2\Phi-\beta\partial\gamma\,.
\end{equation}
The action \eqref{eq:ads full action} may also be obtained from an $\mathfrak{sl}(2,\mathbb{R})_{k_b}$ WZW model. Thus, bosonic string theory on $\rm AdS_3$ possesses $\mathfrak{sl}(2,\mathbb{R})_{k_b}$ currents and these can be written in terms of the Wakimoto fields as
\begin{equation}
\begin{aligned}
\mathcal{J}^+=\beta\,,\quad \mathcal{J}^3=-\frac{1}{Q}\partial\Phi+\beta\gamma\,,\quad \mathcal{J}^-=-\frac{2}{Q}\partial\Phi\,\gamma+\beta\gamma^2-k_b\partial\gamma\,.
\label{eq:currents from Wakimoto fields}
\end{aligned}
\end{equation}
The product of fields is defined by the normal ordering defined in Section 2.2 of \cite{Polchinski:1998rq} which is also reviewed in Appendix C of \cite{Gaberdiel:2022als}. Here, $Q$ and $k_b$ are related by
\begin{equation}
Q=\sqrt{\frac{2}{k_b-2}}\,.
\end{equation}
One can then check that the following OPEs are satisfied
\begin{equation}
\begin{gathered}
\mathcal{J}^3(z)\mathcal{J}^{\pm}(w)\sim\pm\frac{\mathcal{J}^{\pm}(w)}{z-w}\,,\quad \mathcal{J}^+(z)\mathcal{J}^-(w)\sim\frac{k_b}{(z-w)^2}-\frac{2\mathcal{J}^3(w)}{z-w}\,,\\
\mathcal{J}^3(z)\mathcal{J}^3(w)\sim-\frac{k_b/2}{(z-w)^2}\,.
\label{eq:bosonic OPEs}
\end{gathered}
\end{equation}

\subsection{Superstrings}
To describe superstrings in ${\rm AdS}_3$, we promote the worldsheet theory from a purely bosonic theory to an $\mathcal{N}=1$ theory. In particular, the WZW model that describes superstrings in $\rm AdS_3$ is an $\mathcal{N}=1$ $\mathfrak{sl}^{(1)}(2,\mathbb{R})_k$ WZW model. The currents $J^a$ and the fermions $\psi^a$ in the theory satisfy the OPEs
\begin{equation}
\begin{gathered}
J^+(z)J^-(w)=\frac{-2J^3(w)}{z-w}+\frac{k}{(z-w)^2},\quad J^3(z)J^\pm(w)= \frac{\pm J^\pm(w)}{z-w},\quad\\
J^3(z)J^3(w)=\frac{-k}{2(z-w)^2}\,,
\label{eq: susy OPEs 1}
\end{gathered}
\end{equation}
for the current-current OPEs and\footnote{Note that here, we have rescaled the fermions $\psi^a$ relative to \cite{Ferreira:2017pgt,Sriprachyakul:2024gyl} so that the fermion OPEs take the canonical form.} 
\begin{equation}
\begin{gathered}
J^\pm(z)\psi^3(w)=\mp \frac{\psi^\pm}{z-w},\quad J^3(z)\psi^\pm(w)=\pm \frac{\psi^\pm}{z-w},\quad J^\pm(z)\psi^\mp(w)=\mp\frac{2\psi^3}{z-w},\\
\psi^+(z)\psi^-(w)=\frac{1}{z-w},\quad \psi^3(z)\psi^3(w)=-\frac{1}{2(z-w)}\,,
\label{eq: susy OPEs 2}
\end{gathered}
\end{equation}
for the current-fermion and fermion-fermion OPEs.
A convenient way to treat superstrings in $\rm AdS_3$ is to decouple the currents and the fermions. Indeed, one may check that the following currents have trivial OPEs with the fermions $\psi^a$ and satisfy the $\mathfrak{sl}(2,\mathbb{R})$ Kac-Moody algebra with level $k+2$
\begin{equation}
\begin{gathered}
\mathcal{J}^+:=J^++2\psi^3\psi^+\,,\quad\mathcal{J}^-:=J^--2\psi^3\psi^-\,,\quad\mathcal{J}^3:=J^3+\psi^-\psi^+\,.
\label{eq:definition of decoupled currents}
\end{gathered}
\end{equation}
Since we restrict ourselves to worldsheets that live near the conformal boundary of $\rm AdS_3$, we can express the currents $\mathcal{J}^a$ in terms of the Wakimoto fields as in \eqref{eq:currents from Wakimoto fields} with an identification $k_b=k+2$. In particular, we now have that
\begin{equation}
Q=\sqrt{\frac{2}{k}}\,.
\end{equation}
The stress tensor and the supercurrent in this case are given by
\begin{equation}
\begin{aligned}
T_{\rm AdS_3}=&-\frac{1}{2}\partial\Phi\partial\Phi-\frac{Q}{2}\partial^2\Phi-\beta\partial\gamma-\frac{1}{2}\left( \psi^+\partial\psi^-+\psi^-\partial\psi^+ \right)+\psi^3\partial\psi^3\,,\\
G_{\rm AdS_3}=&\frac{1}{\sqrt{k}}\left( \mathcal{J}^+\psi^-+\mathcal{J}^-\psi^+-2\mathcal{J}^3\psi^3-2\psi^+\psi^-\psi^3 \right)\\
=&\frac{1}{\sqrt{k}}\left( \beta\psi^-_\gamma+\frac{2\partial\Phi}{Q}\psi^3_\gamma-(k+2)\partial\gamma\psi^+-2\psi^+\psi^-\psi^3 \right)\,.
\end{aligned}
\end{equation}
In writing the last line, we have used the definition \eqref{eq:definition of superpartner fermions} and the expressions \eqref{eq:currents from Wakimoto fields} for $\mathcal{J}^a$. These fields satisfy the $\mathcal{N}=1$ OPEs, that is 
\begin{equation}
\begin{aligned}
T_{\rm AdS_3}(z)T_{\rm AdS_3}(w)\sim&\frac{c_{\rm AdS_3}}{2(z-w)^4}+\frac{2T_{\rm AdS_3}(w)}{(z-w)^2}+\frac{\partial_wT_{\rm AdS_3}}{z-w}\,,\\
T_{\rm AdS_3}(z)G_{\rm AdS_3}(w)\sim&\frac{3G_{\rm AdS_3}(w)}{2(z-w)^2}+\frac{\partial_wG_{\rm AdS_3}}{z-w}\,,\\
G_{\rm AdS_3}(z)G_{\rm AdS_3}(w)\sim&\frac{2c_{\rm AdS_3}}{3(z-w)^3}+\frac{2T_{\rm AdS_3}(w)}{z-w}\,.
\label{eq:N=1 OPEs}
\end{aligned}
\end{equation}

\subsection{The dual CFT}
It was argued in \cite{Eberhardt:2021vsx,Dei:2022pkr,Knighton:2023mhq,Knighton:2024qxd} that the dual CFT to strings in ${\rm AdS}_3\times X$ is
\begin{equation}
\text{Sym}^N(\mathbb{R}_{\widetilde{Q}}\times X)+\mu\int\sigma_{2,\alpha}\,,
\end{equation}
for bosonic strings and it was also argued in \cite{Eberhardt:2021vsx,Sriprachyakul:2024gyl} that the dual CFT should be\footnote{In \cite{Eberhardt:2021vsx}, the compact direction was taken to be ${\rm S^3}\times \mathbb{T}^4$ or ${\rm S^3} \times K3$. \cite{Sriprachyakul:2024gyl} then generalises the proposal to any compact space $X$ that is a consistent string background. Some discussions that concern consistent string vacua in other compact manifolds $X$ can be found, for example, in \cite{Balthazar:2021xeh,Martinec:2021vpk,Berenstein:1999gj,Giveon:1999jg}.}
\begin{equation}
\text{Sym}^N(\mathbb{R}^{(1)}_{\widetilde{Q}}\times X)+\mu\int\sigma_{2,X,\alpha}\,.
\end{equation}
for superstrings. Here, $\mathbb{R}_{\widetilde{Q}}$ denotes a linear dilaton theory with dilaton slope $\widetilde{Q}$ and the superscript $(1)$ means that the dilaton theory is now an $\mathcal{N}=1$ supersymmetric linear dilaton theory. The spacetime dilaton slope $\widetilde{Q}$ is related to the worldsheet dilaton slope $Q$ via
\begin{equation}
\widetilde{Q}=Q-\frac{2}{Q}\,.
\end{equation}
$\sigma_2$ denotes a twist-2 marginal deformation and the subscript $\alpha$ denotes the dilaton momentum of the deformation which is
\begin{equation}
\alpha=\frac{1}{Q}\,.
\end{equation}
In the supersymmetric case, we also put an additional subscript $X$ to emphasise that the precise form of the deformation could depend on the compact manifold whereas in the bosonic string, the deformation is independent of $X$ \cite{Knighton:2024qxd,Eberhardt:2021vsx}. The main evidence comes from comparing the spectra and ground state correlation functions on both sides of the duality. However, since the dual CFT is conjectured to contain a linear dilaton theory, it should be possible to construct a DDF operator corresponding to this dilaton theory from the worldsheet and this is the task of the next section.

\section{DDF operators}\label{section:DDF construction}
In this section, we construct the DDF operator for the dilaton theory $\mathbb{R}_{\widetilde{Q}}$ and show that it reproduces expected commutation relations with itself and with the spacetime Virasoro generators.

\subsection{Bosonic strings}\label{subsection:bosonic DDF}
Motivated by the discussion in Section 5 of \cite{Knighton:2024qxd}, we make an ansatz of the DDF operator as follows\footnote{Here, by $\oint dz$ we mean $\frac{1}{2\pi i}\oint_0dz$, unless stated otherwise.}
\begin{equation}
\varphi^{bos}_n:=\oint dz\, \gamma^n\left( \partial\Phi+a\frac{\partial^2\gamma}{\partial\gamma} \right)\,.
\label{eq:bosonic DDF ansatz}
\end{equation}
The constant $a$ is determined by requiring the above operator to be physical, that is, to commute with the BRST charge. This condition is fulfilled if the integrand in \eqref{eq:bosonic DDF ansatz} is a conformal primary of weight 1, see Appendix \ref{appendix:BRST analysis}. In the bosonic string theory, the BRST charge is given by
\begin{equation}
Q_{BRST}:=\oint dz\,c\left( T^m+\frac{1}{2}T^{gh} \right)
\end{equation}
where $T^m$ is the matter stress tensor and $T^{gh}$ is the stress tensor for the $bc$ conformal ghosts. Explicitly, we have
\begin{equation}
T^m:=-\frac{1}{2}\partial\Phi\partial\Phi-\frac{Q}{2}\partial^2\Phi-\beta\partial\gamma+T^X\,.
\end{equation}
Thus, requiring the integrand in \eqref{eq:bosonic DDF ansatz} to be a conformal primary of weight 1 implies, in particular, that the third order pole vanishes which results in 
\begin{equation}
a=\frac{Q}{2}\,.
\end{equation}
That is, we now have
\begin{equation}
\begin{aligned}
\varphi^{bos}_n:=\oint dz\, \gamma^n\left( \partial\Phi+\frac{Q}{2}\frac{\partial^2\gamma}{\partial\gamma} \right)\,.
\label{eq:bosonic DDF for phi}
\end{aligned}
\end{equation}
As \eqref{eq:bosonic DDF for phi} is expected to correspond to modes of the spacetime dilaton field, it should satisfy the free boson commutation relation and it should have the correct commutation relation with the spacetime Virasoro generators as we now show.

Let us first consider the commutator of \eqref{eq:bosonic DDF for phi} with itself, we have 
\begin{equation}
\begin{aligned}
[\varphi^{bos}_m,\varphi^{bos}_n]=&\oint_0dw\oint_wdz\,\gamma^m(z)\left( \partial\Phi+\frac{Q}{2}\frac{\partial^2\gamma}{\partial\gamma} \right)(z)\gamma^n(w)\left( \partial\Phi+\frac{Q}{2}\frac{\partial^2\gamma}{\partial\gamma} \right)(w)\\
=&\oint_0dw\oint_wdz\,\gamma^m(z)\gamma^n(w)\left( -\frac{1}{(z-w)^2} \right)(w)\\
=&-m\oint dw\,\gamma^{m+n-1}\partial\gamma\\
=&-m\delta_{m+n,0}{\mathcal{I}}\,,
\label{eq:Phi Phi OPE in bosonic strings}
\end{aligned}
\end{equation}
which we see agreeing with the commutation relation for the modes of a free boson $\partial\varphi$. We have also denoted by $\mathcal{I}$ the identity operator
\begin{equation}
\mathcal{I}:=\oint dw\, \gamma^{-1}\partial\gamma\,.
\end{equation}
Next, the spacetime Virasoro generators are given by \cite{Giveon:1998ns,Eberhardt:2019qcl}\footnote{The spacetime Virasoro generators were first discovered in \cite{Giveon:1998ns} but our convention is that of \cite{Eberhardt:2019qcl}.}
\begin{equation}
\begin{aligned}
\mathcal{L}_m:=&\oint dz \left( (1-m^2)\gamma^m\mathcal{J}^3+\frac{m(m-1)}{2}\gamma^{m+1}\mathcal{J}^++\frac{m(m+1)}{2}\gamma^{m-1}\mathcal{J}^- \right)\\
=&\oint dz\left( \beta\gamma^{m+1}-\frac{(m+1)\gamma^m\partial\Phi}{Q} \right)\,,
\end{aligned}
\end{equation}
where in the second line we have substituted in the expressions \eqref{eq:currents from Wakimoto fields} and we have discarded the term $\oint dz n\gamma^{n-1}\partial\gamma=n\delta_{n,0}\mathcal{I}=0$. Using this and \eqref{eq:bosonic DDF for phi}, we see that
\begin{equation}
\begin{aligned}
[\mathcal{L}_m,\varphi^{bos}_n]=-n\varphi^{bos}_{m+n}-\frac{\widetilde{Q}m(m+1)}{2}\delta_{m+n,0}\mathcal{I}\,.
\label{eq:Vir commu w/ bosonic dilaton}
\end{aligned}
\end{equation}
We perform this computation in Appendix \ref{appendix:L phi commu}.

Note that if we keep the coefficient $a$ in \eqref{eq:bosonic DDF ansatz} generic and compute $[\mathcal{L}_m,\varphi^{bos}_n]$, we would instead obtain
\begin{equation}
\widetilde{Q}_a=2a-\frac{2}{Q}
\end{equation}
as the spacetime dilaton slope. Hence, we see that the value of $a$ that was fixed by the BRST condition also automatically gives the correct dilaton slope as predicted by the correlation function computations \cite{Eberhardt:2021vsx,Knighton:2024qxd}. Incidentally, the integrand of \eqref{eq:bosonic DDF for phi} coincides, up to the conformal factor $(\partial\gamma)^{-1}$, with the conformal transformation of $\Phi$, see eq.(C.2) of \cite{Eberhardt:2019ywk}. Note also the different normalisation $\sqrt{2k}\Phi_{\rm there}=\Phi_{\rm here}$ which comes from the OPE in eq.(6.2) in the same paper. Thus, one could have expected the result \eqref{eq:bosonic DDF for phi} purely on the ground of conformal transformation property of $\Phi$ and we have shown that it leads indeed to the right answer.

\subsection{Superstrings}
We now consider the supersymmetric version of \eqref{eq:bosonic DDF for phi}. Unfortunately, the expression \eqref{eq:bosonic DDF for phi} does not continue to commute with the superstring BRST charge. In particular, it does not commute with the supercurrent part of the BRST operator. Note that the BRST charge is now given by
\begin{equation}
Q_{BRST}:=\oint dz\left(c\left( T^m+\frac{1}{2}T^{gh} \right)+\hat{\gamma}\left( G^m+\frac{1}{2}G^{gh} \right)\right)\,,
\label{eq:susic BRST charge}
\end{equation}
where $\hat{\gamma}$, together with $\hat{\beta}$, form the superconformal ghost $\hat{\beta}\hat{\gamma}$ system. To construct a supersymmetric analogue of \eqref{eq:bosonic DDF for phi}, we note two observations that will guide us towards to the answer. Firstly, it can be shown that an operator constructed from contour integration of $G_{-\frac{1}{2}}$ descendant of a (matter) superconformal primary of weight $\tfrac{1}{2}$ commutes with the BRST operator. We give a derivation of this simple fact in Appendix \ref{appendix:BRST analysis}. Secondly, the superconformal primary of weight $\tfrac{1}{2}$ should be constructed from the $G_{\frac{1}{2}}$ descendant of $\partial\Phi+\tfrac{Q}{2}\tfrac{\partial^2\gamma}{\partial\gamma}$. The second observation is quite implicit in the discussion around equations (3.4-3.5) of \cite{Giveon:1998ns} but we will see that it leads to the desired result. 

The $G_{\frac{1}{2}}$ descendants can be extracted from the coefficients of the second order poles of the $G(z)\partial_w\Phi$ and $G(z)\tfrac{\partial_w^2\gamma}{\partial_w\gamma}$ OPEs. The results are
\begin{equation}
\begin{aligned}
G_{\frac{1}{2}}\cdot\partial\Phi=&-\frac{2}{Q}\psi^3_\gamma\,,\\
G_{\frac{1}{2}}\cdot\frac{\partial^2\gamma}{\partial\gamma}=&\frac{Q}{2}\left( \frac{\partial^2\gamma\psi^-_\gamma}{(\partial\gamma)^2}-\frac{2\partial\psi^-_\gamma}{\partial\gamma} \right)\,.
\end{aligned}
\end{equation}
Here, we have defined the shorthand
\begin{equation}
\begin{aligned}
\psi^3_\gamma:=&\psi^3-\gamma\psi^+\,,\\
\psi^-_\gamma:=&\psi^--2\gamma\psi^3+\gamma^2\psi^+\,.
\label{eq:definition of superpartner fermions}
\end{aligned}
\end{equation}
Actually, the fermions defined above are superpartners of $\Phi$ and $\gamma$, up to some normalisation, respectively. Hence, our ansatz for the dilaton DDF operator for superstrings is
\begin{equation}
\begin{aligned}
\varphi^{susy}_n=&\frac{1}{\sqrt{k}}\oint dz\,G_{-\frac{1}{2}}\cdot\left(\gamma^nG_{\frac{1}{2}}\cdot\left( \partial\Phi+\frac{Q}{2}\frac{\partial^2\gamma}{\partial\gamma} \right)\right)\\
=&-\sqrt{2}\oint dz\,G_{-\frac{1}{2}}\cdot\left(\gamma^n\psi^3_\gamma-\frac{\gamma^n}{2k}\left( \frac{\psi^-_\gamma\partial^2\gamma}{(\partial\gamma)^2}-\frac{2\partial\psi^-_\gamma}{\partial\gamma} \right) \right)\,.
\label{eq:Phi DDF for superstrings, umsimplified}
\end{aligned}
\end{equation}
The prefactor is for later convenience and it will turn out to be the right normalisation which reproduces \eqref{eq:Phi Phi OPE in bosonic strings} for superstrings. To save some writing, let us define
\begin{equation}
\begin{aligned}
A(z):=&\gamma^n\psi^3_\gamma\,,\\
B(z):=&\gamma^n\frac{\partial\psi^-_\gamma}{\partial\gamma}\,,\\
C(z):=&\frac{\gamma^n\partial^2\gamma\psi^-_\gamma}{(\partial\gamma)^2}\,.
\label{eq:definition of ABC}
\end{aligned}
\end{equation}
The OPEs between $G(z)$ and the fields defined above can be found by the usual Wick contraction and this gives
\begin{equation}
\begin{aligned}
G(z)A(w)=&\frac{1}{\sqrt{k}}\left( \frac{\gamma^n(w)}{(z-w)^2}-\frac{\frac{\partial\Phi\gamma^n}{Q}+\partial\gamma^n+n\gamma^{n-1}\psi^-_\gamma\psi^3_\gamma}{(z-w)}\right)\,,\\
G(z)B(w)=&\frac{1}{\sqrt{k}}\left( \frac{-(k+1)\gamma^n(w)+\gamma^n\frac{\psi^-_\gamma\partial\psi^-_\gamma}{(\partial\gamma)^2}}{(z-w)^2}+\frac{-k\gamma^n\frac{\partial^2\gamma}{\partial\gamma}+\partial\gamma^n-\frac{n\gamma^{n-1}\psi^-_\gamma\partial\psi^-_\gamma}{\partial\gamma}}{(z-w)}\right)\,,\\
G(z)C(w)=&\frac{1}{\sqrt{k}}\left( \frac{-2\gamma^n(w)+2\gamma^n\frac{\psi^-_\gamma\partial\psi^-_\gamma}{(\partial\gamma)^2}}{(z-w)^2}+\frac{-k\gamma^n\frac{\partial^2\gamma}{\partial\gamma}-\frac{2\gamma^{n}\psi^-_\gamma\partial\psi^-_\gamma\partial^2\gamma}{(\partial\gamma)^3}+\frac{\gamma^n\psi^-_\gamma\partial^2\psi^-_\gamma}{(\partial\gamma)^2}}{(z-w)}\right)\,.
\end{aligned}
\end{equation}
From the OPEs above, we see that the combination
\begin{equation}
\gamma^n\psi^3_\gamma-\frac{\gamma^n}{2k}\left( \frac{\psi^-_\gamma\partial^2\gamma}{(\partial\gamma)^2}-\frac{2\partial\psi^-_\gamma}{\partial\gamma} \right)
\end{equation}
is a superconformal primary, i.e., its OPE with $G$ only has a simple pole singularity and its OPE with $T$ has at most a double pole singularity.
Thus, \eqref{eq:Phi DDF for superstrings, umsimplified} becomes
\begin{equation}
\begin{aligned}
\varphi^{susy}_n=&\oint dz\, \gamma^n\left( \partial\Phi+\frac{Q}{2}\frac{\partial^2\gamma}{\partial\gamma}+\frac{nQ\psi^-_\gamma\psi^3_\gamma}{\gamma}+\frac{Q\partial\gamma^n\psi^-_\gamma\partial\psi^-_\gamma}{k\gamma^n(\partial\gamma)^2}+\frac{Q}{2k}\partial\left(\frac{\psi^-_\gamma\partial\psi^-_\gamma}{(\partial\gamma)^2}\right)\right)\\
=&\oint dz\, \gamma^n\left( \partial\Phi+\frac{Q}{2}\frac{\partial^2\gamma}{\partial\gamma}+\frac{nQ\psi^-_\gamma\psi^3_\gamma}{\gamma}+\frac{Q\partial\gamma^n\psi^-_\gamma\partial\psi^-_\gamma}{2k\gamma^n(\partial\gamma)^2}\right)\\
=&\oint dz\, \gamma^n\left( \partial\Phi+\frac{Q}{2}\frac{\partial^2\gamma}{\partial\gamma}+\frac{nQ\psi^-_\gamma\psi^3_\gamma}{\gamma}+\frac{nQ\psi^-_\gamma\partial\psi^-_\gamma}{2k\gamma\partial\gamma}\right)\,.
\label{eq:Phi DDF for superstrings, simplified}
\end{aligned}
\end{equation}
In going to the second line, we have performed the integration by parts and discarded the total derivative term. It is not hard to see that \eqref{eq:Phi Phi OPE in bosonic strings} still holds as we now show. First, we note some useful Wick contractions
\begin{equation}
\begin{aligned}
\overbracket{\psi^-_\gamma(z)\psi^-_\gamma}(w)=&\frac{(\gamma(z)-\gamma(w))^2}{z-w}\,,\\
\overbracket{\psi^3_\gamma(z)\psi^3_\gamma}(w)=&-\frac{1}{2}\frac{1}{z-w}\,,\\
\overbracket{\psi^-_\gamma(z)\psi^3_\gamma}(w)=&\frac{\gamma(z)-\gamma(w)}{z-w}\,.
\label{eq:useful contractions}
\end{aligned}
\end{equation}
Thus, the following OPEs are all regular since
\begin{equation}
\begin{aligned}
\left( \psi^-_\gamma\psi^3_\gamma \right)(z)\left( \psi^-_\gamma\psi^3_\gamma \right)(w)\sim&-\frac{\psi^3_\gamma(z)\psi^3_\gamma(w)(\gamma(z)-\gamma(w))^2}{z-w}+\frac{\psi^3_\gamma(z)\psi^-_\gamma(w)(\gamma(z)-\gamma(w))}{z-w}\\
&-\frac{\psi^-_\gamma(z)\psi^3_\gamma(w)(\gamma(z)-\gamma(w))}{z-w}+\frac{1}{2}\frac{\psi^-_\gamma(z)\psi^-_\gamma(w)}{z-w}\\
&-\frac{(\gamma(z)-\gamma(w))^2}{(z-w)^2}+\frac{1}{2}\frac{(\gamma(z)-\gamma(w))^2}{(z-w)^2}\\
\sim&O(1)\,,
\end{aligned}
\end{equation}
\begin{equation}
\begin{aligned}
\left( \psi^-_\gamma\psi^3_\gamma \right)(z)\left( \psi^-_\gamma\partial\psi^-_\gamma \right)(w)\sim&-\frac{\psi^3_\gamma(z)\partial_w\psi^-_\gamma(\gamma(z)-\gamma(w))^2}{z-w}+\psi^3_\gamma(z)\psi^-_\gamma\partial_w\left(\frac{(\gamma(z)-\gamma(w))^2}{z-w}\right)\\
&-\frac{\psi^-_\gamma(z)\partial_w\psi^-_\gamma(\gamma(z)-\gamma(w))}{z-w}+\psi^-_\gamma(z)\psi^-_\gamma(w)\partial_w\left(\frac{\gamma(z)-\gamma(w)}{z-w}\right)\\
&+\frac{(\gamma(z)-\gamma(w))^2}{z-w}\partial_w\left(\frac{\gamma(z)-\gamma(w)}{z-w}\right)\\
&-\frac{\gamma(z)-\gamma(w)}{z-w}\partial_w\left(\frac{(\gamma(z)-\gamma(w))^2}{z-w}\right)\\
\sim&O(1)\,,
\end{aligned}
\end{equation}
and
\begin{equation}
\begin{aligned}
&\hspace{-2cm}\left( \psi^-_\gamma\partial\psi^-_\gamma \right)(z)\left( \psi^-_\gamma\partial\psi^-_\gamma \right)(w)\\
\hspace{1cm}\sim&-\frac{(\gamma(z)-\gamma(w))^2\partial_z\psi^-_\gamma\partial_w\psi^-_\gamma}{z-w}+\partial_z\psi^-_\gamma\psi^-_\gamma(w)\partial_w\left(\frac{(\gamma(z)-\gamma(w))^2}{z-w}\right)\\
&+\psi^-_\gamma(z)\partial_w\psi^-_\gamma\partial_z\left(\frac{(\gamma(z)-\gamma(w))^2}{z-w}\right)-\psi^-_\gamma(z)\psi^-_\gamma(w)\partial_z\partial_w\left( 
\frac{(\gamma(z)-\gamma(w))^2}{z-w} \right)\\
&-\frac{(\gamma(z)-\gamma(w))^2}{z-w}\partial_z\partial_w\left( 
\frac{(\gamma(z)-\gamma(w))^2}{z-w} \right)\\
&+\partial_z\left( \frac{(\gamma(z)-\gamma(w))^2}{z-w}\right)\partial_w\left( \frac{(\gamma(z)-\gamma(w))^2}{z-w} \right)\\
\sim& O(1)\,.
\end{aligned}
\end{equation}
Therefore, 
\begin{equation}
\begin{aligned}
[\varphi^{susy}_m,\varphi^{susy}_n]=&\oint_0dw\oint_wdz\,\gamma^m(z)\partial_z\Phi\gamma^n(w)\partial_w\Phi\\
=&-m\delta_{m+n,0}\mathcal{I}\,,
\end{aligned}
\end{equation}
where we only keep the term with nontrivial OPE in the first line. Furthermore, to see that \eqref{eq:Phi DDF for superstrings, simplified} indeed corresponds to the linear dilaton, we compute the commutation relation $[\mathcal{L}^{susy}_m,\varphi^{susy}_n]$. $\mathcal{L}^{susy}_m$ is the spacetime  Virasoro generators in superstrings and is given by \cite{Giveon:1998ns,Eberhardt:2019qcl}
\begin{equation}
\begin{aligned}
&\hspace{-0.8cm}\mathcal{L}^{susy}_m\\
:=&\oint dz \left( (1-m^2)\gamma^m{J}^3+\frac{m(m-1)}{2}\gamma^{m+1}{J}^++\frac{m(m+1)}{2}\gamma^{m-1}{J}^- \right)\\
=&\oint dz\left( \beta\gamma^{m+1}-\frac{(m+1)\gamma^m\partial\Phi}{Q}+(m^2+m)\gamma^{m-1}\psi^3_\gamma\psi^-_\gamma+m\gamma^m\psi^+\psi^-_\gamma+\gamma^m\psi^+\psi^-\right)\,.
\end{aligned}
\end{equation}
In going to the second line, we used \eqref{eq:currents from Wakimoto fields}, \eqref{eq:definition of decoupled currents} and \eqref{eq:definition of superpartner fermions}. The commutation relation is calculated in Appendix \ref{appendix:L phi OPE in superstrings} and the result is
\begin{equation}
[\mathcal{L}^{susy}_m,\varphi^{susy}_n]=-n\varphi^{susy}_{m+n}-\frac{\widetilde{Q}m(m+1)}{2}\delta_{m+n,0}\mathcal{I}\,.
\label{eq:Vir commu w/ susy dilaton}
\end{equation}
Thus, we conclude that \eqref{eq:Phi DDF for superstrings, simplified} is indeed corresponding to the spacetime linear dilaton. We discuss the consequences of our result in the next section.

\section{Discussion}\label{seciton:discussion}
Let us begin by summarising our results. We have constructed worldsheet DDF operators, see equations \eqref{eq:bosonic DDF for phi} and \eqref{eq:Phi DDF for superstrings, simplified}, that correspond to the spacetime linear dilaton field in bosonic string theory and superstring theory respectively. These DDF operators commute with the BRST charge and thus, map physical states to physical states. The operator \eqref{eq:Phi DDF for superstrings, simplified} also has an even number of fermions which means that it commutes with the fermion number operator $F, \bar F$ and thus, it respects the GSO projection. Our results tell us the precise correspondence between dilaton excited states on the dual CFT and the worldsheet states. The existence of \eqref{eq:bosonic DDF for phi} and \eqref{eq:Phi DDF for superstrings, simplified} and the commutation relations \eqref{eq:Vir commu w/ bosonic dilaton}, \eqref{eq:Vir commu w/ susy dilaton} also tell us that the seed theory indeed contains a linear dilaton factor and the fact that these DDF operators are constructed purely from the $\rm AdS_3$ fields means that the dilaton theory appears generically in $\rm AdS_3/CFT_2$ holography with pure NSNS flux. Hence, this further supports the proposal in \cite{Sriprachyakul:2024gyl} which was made by looking only at some spectrally-flowed ground state correlation functions. Although we did not explicitly discuss the possible values of the mode number $n$ in \eqref{eq:bosonic DDF for phi}, \eqref{eq:Phi DDF for superstrings, simplified}, we expect that the discussion of Sections (2.4-2.5) in \cite{Eberhardt:2019qcl} will go through and thus that the mode number $n$ is appropriately fractional for $w$-spectrally flowed states.

There are some curious points which we would like to mention here which may be interesting. First of all, eq.(5.5) of \cite{Knighton:2024qxd} tells us that one may expect the equation \eqref{eq:bosonic DDF for phi} to instead be
\begin{equation}
\varphi_n\overset{!}{=}\oint dz\,\gamma^n\left(\partial\Phi+\frac{1}{Q}\frac{\partial^2\gamma}{\partial\gamma}\right)\,.
\end{equation}
However, as we saw in Section \ref{subsection:bosonic DDF}, this does not commute with the BRST charge nor does it lead to the correct commutation relation with the spacetime Virasoro generators. Secondly, one may ask what happens to the operators \eqref{eq:bosonic DDF for phi}, \eqref{eq:Phi DDF for superstrings, simplified} when $k=1$ (or, equivalently, $k_b=3$) and $X={\rm S^3}\times \mathbb{T}^4$? As discussed in \cite{Eberhardt:2021vsx}, we expect the factor $\mathbb{R}_{\widetilde{Q}=0}\times \mathfrak{su}(2)^{(1)}_1$ in the seed theory of the dual CFT to become trivial. However, we do not currently know how such a mechanism manifests itself on the level of DDF operators. That being said, we do not expect the operators \eqref{eq:bosonic DDF for phi}, \eqref{eq:Phi DDF for superstrings, simplified} to become BRST exact whenever $k=1$. For one, being BRST exact should mean that the spacetime dilaton simply disappears from the dual CFT, but since the central charge of the spacetime dilaton does not vanish at $k=1$ (it is, in fact, equal to 1), the linear dilaton theory should not be a trivial theory on its own. For another, there is a background\footnote{One such example is superstrings in ${\rm AdS}_3\times {\rm S}^3_{k^+}\times {\rm S}^3_{k^-}\times {\rm S}^1$, the subscripts denote the fluxes through the 3-spheres. For $k^+=k^-=2$, the level of $\mathfrak{sl}(2,\mathbb{R})$ Kac-Moody algebra is $k=1$. However, there is no unitarity problem in describing $\mathfrak{su}(2)^{(1)}_2\times\mathfrak{su}(2)^{(1)}_2$ WZW model in the RNS formalism. Indeed, $\mathfrak{su}(2)^{(1)}_2$ decomposes into the bosonic $\mathfrak{su}(2)_{2-2=0}$ WZW model, which is a trivial theory and 3 free fermions.} in which $k=1$ but the compact manifold description does not suffer from the unitarity problem as $\mathfrak{su}(2)^{(1)}_1$ does. In this case, we do not see reasons why the dilaton should disappear and why the corresponding operators should be BRST exact.

Moreover, note that \eqref{eq:Phi Phi OPE in bosonic strings} seems to imply that the spacetime dilaton is free. However, one could be worried that this may be in conflict with the presence of the CFT deformation term $\int\sigma_{2,\alpha,X}$. We currently do not have a complete understanding as to why the dilaton seems to be free but we provide an intuitive, and somewhat heuristic, argument as to why the two facts should not conflict each other. We suspect that since the deformation $\int\sigma_{2,\alpha,X}$ is treated only perturbatively in the dual CFT, the spacetime dilaton $\varphi$ thus satisfies the free boson OPE. This then results in \eqref{eq:Phi Phi OPE in bosonic strings} since the near-boundary approximation only captures the perturbative aspects of the dual CFT.

\acknowledgments
I thank Matthias Gaberdiel for detailed and stimulating discussion, Kiarash Naderi and Beat Nairz for discussions. I also thank Matthias Gaberdiel and Beat Nairz for carefully reading and comments on the early version of the paper. The work of VS is supported by a grant from the Swiss National Science Foundation. The activities of the group are more generally supported by the NCCR SwissMAP, which is also funded by the Swiss National Science Foundation. 

\appendix
\section{Assorted computations}\label{appendix:assorted calculations}

\subsection{BRST analysis}\label{appendix:BRST analysis}
Let us first start with bosonic strings. We begin by showing that
\begin{equation}
\mathcal{A}:=\oint dz\, A(z)
\end{equation}
commutes with $Q_{BRST}$ if $A(z)$ is a conformal matter field of weight 1. Since $A(z)$ is a matter field, it has trivial OPE with the purely ghost part of $Q_{BRST}$, hence, we only need to consider the $T^m$ part. $A(z)$ is conformal primary of weight 1 means
\begin{equation}
T^m(z)A(w)\sim\frac{A(w)}{(z-w)^2}+\frac{\partial_wA}{z-w}\,.
\end{equation}
Thus, we have
\begin{equation}
\begin{aligned}
[Q_{BRST},\mathcal{A}]=&\oint_0dw\oint_wdz\,c(z)T^m(z)A(w)\\
=&\oint_0dw\oint_wdz\,c(z)\left( \frac{A(w)}{(z-w)^2}+\frac{\partial_wA}{z-w} \right)\\
=&\oint_0dw\left( \partial_wcA(w)+c(w)\partial_wA \right)\\
=&\oint_0 dw\,\partial_w(cA)\\
=&0\,.
\end{aligned}
\end{equation}
Hence, $\mathcal{A}$ commutes with $Q_{BRST}$ and $\mathcal{A}$ is therefore physical. We are now ready to consider a similar problem in superstrings. Let us assume that $B(z)$\footnote{We emphasise that the fields $A(z), B(z)$ in this appendix are not related to the definition \eqref{eq:definition of ABC}. We also implicitly assume that $B(z)$ is in the NS sector so that $G(z)$ is half-integer moded.} is a superconformal matter field of weight $\tfrac{1}{2}$, that is, we have
\begin{equation}
\begin{aligned}
G^m(z)B(w)\sim&\frac{(G^m_{-\frac{1}{2}}\cdot B)(w)}{z-w}\,,\\
T^m(z)B(w)\sim&\frac{1}{2}\frac{B(w)}{(z-w)^2}+\frac{\partial_wB}{z-w}\,,
\end{aligned}
\end{equation}
and that the OPEs between $B(z)$ and the ghosts $b,c,\hat{\beta},\hat{\gamma}$ are all trivial. Define $A(z):=(G^m_{-\frac{1}{2}}\cdot B)(z)$, we see that $A(z)$ is a conformal primary matter field of weight 1, however, $A(z)$ is not a superconformal primary since
\begin{equation}
\begin{aligned}
G^m_{\frac{1}{2}}\cdot A=&G^m_{\frac{1}{2}}\cdot\left(G^m_{-\frac{1}{2}}\cdot B\right)\\
=&\left( \{G^m_{-\frac{1}{2}},G^m_{\frac{1}{2}}\} \right)\cdot B\\
=&2L^m_0\cdot B\\
=&B\,.
\end{aligned}
\end{equation}
In computing the anticomutator, we have used the appropriate OPE \eqref{eq:N=1 OPEs} (with suitable modifications to account for the $X$ part). Also, since we have
\begin{equation}
\begin{aligned}
G^m_{-\frac{1}{2}}\cdot\left(G^m_{-\frac{1}{2}}\cdot B\right)=&\frac{1}{2}\left( \{G^m_{-\frac{1}{2}},G^m_{-\frac{1}{2}}\} \right)\cdot B\\
=&L^m_{-1}\cdot B\\
=&\partial B\,,
\end{aligned}
\end{equation}
the $G^m(z)A(w)$ OPE takes the following form
\begin{equation}
G^m(z)A(w)\sim\frac{B(w)}{(z-w)^2}+\frac{\partial_wB}{z-w}\,.
\end{equation}
Thus, we now have
\begin{equation}
\begin{aligned}
[Q_{BRST},\mathcal{A}]_{\pm}=&\oint_0dw\oint_wdz\Bigl(c(z)T^m(z)+\hat{\gamma}(z)G^m(z)\Bigr)A(w)\\
=&\oint_0dw\oint_wdz\Biggl(c(z)\left( \frac{A(w)}{(z-w)^2}+\frac{\partial_wA}{z-w} \right)+\hat{\gamma}(z)\left( \frac{B(w)}{(z-w)^2}+\frac{\partial_wB}{z-w} \right)\Biggr)\\
=&\oint_0dw\Bigl( \partial_wcA(w)+c(w)\partial_wA +\partial_w\hat{\gamma}B(w)+\hat{\gamma}(w)\partial_wB\Bigr)\\
=&\oint_0 dw\Bigl(\partial_w(cA)+\partial_w(\hat{\gamma}B)\Bigr)\\
=&0\,.
\end{aligned}
\end{equation}
Hence, we see that $Q_{BRST}$ commutes (or anticommutes depending on the statistics of $\mathcal{A}$) with $\mathcal{A}$ as desired.

\subsection[Computating \texorpdfstring{$[\mathcal{L}_m,\varphi^{bos}_n]$}{[mth mode of L,nth mode of bosonic phi]}]{\boldmath Computating \texorpdfstring{$[\mathcal{L}_m,\varphi^{bos}_n]$}{[mth mode of L,nth mode of bosonic phi]}}\label{appendix:L phi commu}
We have
\begin{equation}
\begin{aligned}
[\mathcal{L}_m,\varphi^{bos}_n]=&\oint_0dw\oint_wdz\left( \beta\gamma^{m+1}-\frac{(m+1)\gamma^m\partial\Phi}{Q} \right)(z)\gamma^n(w)\left( \partial\Phi+\frac{Q}{2}\frac{\partial^2\gamma}{\partial\gamma} \right)(w)\\
=&\oint_0dw\oint_wdz\,\gamma^{m+1}(z)\left( \frac{-n\gamma^{n-1}(w)\partial_w\Phi}{z-w}\right.\\
&\hspace{3cm}+\frac{Q}{2}\left( \frac{-n\gamma^{n-1}(w)\partial^2_w\gamma}{(z-w)\partial_w\gamma}+\frac{-2\gamma^n(w)}{(z-w)^3\partial_w\gamma}+\frac{\gamma^n(w)\partial^2_w\gamma}{(z-w)^2(\partial_w\gamma)^2} \right) \\
&\hspace{4cm}\left.+\frac{(m+1)\gamma^m(z)\gamma^n(w)}{Q(z-w)^2}\right)\\
=&\oint dw\left(-n\gamma^{m+n-1}\partial\Phi-\frac{nQ}{2}\frac{\gamma^{m+n}\partial^2\gamma}{\partial\gamma}-\frac{Q(m+1)}{2}\frac{(\gamma\partial^2\gamma+m(\partial\gamma)^2)\gamma
^{m+n-1}}{\partial\gamma}\right.\\
&\hspace{3cm}\left.+\frac{Q(m+1)}{2}\frac{\gamma^{m+n}\partial^2\gamma}{\partial\gamma}+\frac{(m+1)m\gamma^{m+n-1}\partial\gamma}{Q}\right)\\
=&-n\varphi^{bos}_{m+n}-\frac{\widetilde{Q}m(m+1)}{2}\delta_{m+n,0}\mathcal{I}\,.
\label{}
\end{aligned}
\end{equation}
Recall that we have defined $\widetilde{Q}=Q-\tfrac{2}{Q}$.

\subsection[Computing \texorpdfstring{$[\mathcal{L}^{susy}_m,\varphi^{susy}_n]$}{[mth mode of supersymmetric L,nth mode of supersymmetric phi]}]{\boldmath Computing \texorpdfstring{$[\mathcal{L}^{susy}_m,\varphi^{susy}_n]$}{[mth mode of supersymmetric L,nth mode of supersymmetric phi]}}\label{appendix:L phi OPE in superstrings}
To compute the commutation relation, we first compute the following OPEs. Note that the term $\psi^3_\gamma\psi^-_\gamma$ has a trivial OPE with the integrand of $\varphi^{susy}_n$, hence, the only OPEs we need are
\begin{equation}
\begin{aligned}
&\beta(z)\gamma^{m+1}(z)\left( \gamma^n\partial\Phi+\frac{Q}{2}\frac{\gamma^n\partial^2\gamma}{\partial\gamma}+nQ\gamma^{n-1}\psi^-_\gamma\psi^3_\gamma+\frac{nQ}{2k}\frac{\gamma^{n-1}\psi^-_\gamma\partial\psi^-_\gamma}{\partial\gamma}\right)(w)\\
&\sim\gamma^{m+1}(z)\left[ -\frac{n\gamma^{n-1}(w)\partial_w\Phi}{z-w}-\frac{Q}{2}\frac{n\gamma^{n-1}\partial^2_w\gamma}{\partial_w\gamma(z-w)}-\frac{Q\gamma^n(w)}{\partial_w\gamma(z-w)^3}+\frac{Q}{2}\frac{\gamma^n(w)\partial^2_w\gamma}{(\partial_w\gamma)^2(z-w)^2} \right.\\
&\hspace{3cm}-\frac{n(n-1)Q\gamma^{n-2}(w)\psi^-_\gamma(w)\psi^3_\gamma(w)}{z-w}+\frac{nQ\gamma^{n-1}\psi^-_\gamma(w)\psi^+(w)}{z-w}\\
&\hspace{3.5cm}-\frac{n(n-1)Q\gamma^{n-2}(w)\psi^-_\gamma(w)\partial_w\psi^-_\gamma}{2k\partial_w\gamma(z-w)}+\frac{nQ}{k}\frac{\gamma^{n-1}(w)\psi^3_\gamma(w)\partial_w\psi^-_\gamma}{\partial_w\gamma(z-w)}\\
&\hspace{4cm}\left.+\frac{nQ}{k}\frac{\gamma^{n-1}(w)\psi^-_\gamma(w)}{\partial_w\gamma}\partial_w\left( \frac{\psi^3_\gamma(w)}{z-w}\right)+\frac{nQ}{2k}\frac{\gamma^{n-1}(w)\psi^-_\gamma(w)\partial_w\psi^-_\gamma}{(\partial_w\gamma)^2(z-w)^2}\right]\,,
\end{aligned}
\end{equation}
\begin{equation}
\begin{aligned}
\hspace{-5.6cm}-\frac{(m+1)}{Q}\gamma^m(z)\partial_z\Phi\gamma^n(w)\partial_w\Phi\sim\frac{(m+1)}{Q}\frac{\gamma^m(z)\gamma^n(w)}{(z-w)^2}\,,
\end{aligned}
\end{equation}
\begin{equation}
\begin{aligned}
&\hspace{-0cm}m\gamma^m(z)\psi^+(z)\psi^-_\gamma(z)\left( nQ\gamma^{n-1}\psi^-_\gamma\psi^3_\gamma+\frac{nQ}{2k}\frac{\gamma^{n-1}\psi^-_\gamma\partial\psi^-_\gamma}{\partial\gamma} \right)(w)\\
&\hspace{0.4cm}\sim mnQ\gamma^m(z)\left[ -\frac{\psi^-_\gamma(z)\gamma^{n-1}(w)\psi^3_\gamma(w)}{z-w}-\frac{\gamma^{n-1}(w)(\gamma(z)-\gamma(w))}{(z-w)^2}-\frac{\psi^-_\gamma(z)\gamma^{n-1}(w)\partial_w\psi^-_\gamma}{2k\partial_w\gamma(z-w)} \right.\\
&\hspace{4cm}\left.+\frac{\psi^-_\gamma(z)\gamma^{n-1}(w)\psi^-_\gamma(w)}{2k\partial_w\gamma(z-w)^2}+\frac{\gamma^{n-1}(w)\partial_w\gamma}{k(z-w)}\right]\,,
\end{aligned}
\end{equation}
\begin{equation}
\begin{aligned}
&\gamma^m(z)\psi^+(z)\psi^-(z)\left( nQ\gamma^{n-1}\psi^-_\gamma\psi^3_\gamma+\frac{nQ}{2k}\frac{\gamma^{n-1}\psi^-_\gamma\partial\psi^-_\gamma}{\partial\gamma} \right)(w)\\
&\sim nQ\gamma^m(z)\gamma^n(w)\left[ -\frac{\psi^-(z)\gamma^{-1}(w)\psi^3_\gamma(w)}{z-w}+\frac{\psi^+(z)\gamma(w)\psi^3_\gamma(w)}{z-w}+\frac{\psi^+(z)\psi^-_\gamma(w)}{z-w}+\frac{1}{(z-w)^2}\right]\\
&\hspace{1cm}+\frac{nQ\gamma^m(z)\gamma^{n-1}(w)}{2k\partial_w\gamma}\left[ -\frac{\psi^-(z)\partial_w\psi^-_\gamma}{z-w}+\frac{\psi^-(z)\psi^-_\gamma(w)}{(z-w)^2}+\frac{\psi^+(z)\gamma^2(w)\partial_w\psi^-_\gamma}{z-w}\right.\\
&\hspace{5cm}\left.-\psi^+(z)\psi^-_\gamma(w)\partial_w\left( \frac{\gamma^2(w)}{z-w} \right)-\frac{2\gamma(w)\partial_w\gamma}{(z-w)^2} \right]\,.
\end{aligned}
\end{equation}
Collecting only terms that are of the form $\tfrac{\widetilde{A}(w)}{z-w}$ and performing the contour integration $\oint_wdz$, we obtain
\begin{equation}
\begin{aligned}
&\oint_wdz(...)=-n\gamma^{m+n}\partial\Phi-\frac{nQ}{2}\frac{\gamma^{m+n}\partial^2\gamma}{\partial\gamma}-\frac{(m+1)Q\gamma^{m+n-1}(m(\partial\gamma)^2+\gamma\partial^2\gamma)}{2\partial\gamma}\\
&+\frac{(m+1)Q\gamma^{m+n}\partial^2\gamma}{2\partial\gamma}\textcolor{red}{-n(n-1)Q\gamma^{m+n-1}\psi^-_\gamma\psi^3_\gamma+nQ\gamma^{m+n}\psi^-_\gamma\psi^+}\textcolor{blue}{-\frac{n(n-1)Q\gamma^{m+n-1}\psi^-_\gamma\partial\psi^-_\gamma}{2k\partial\gamma}}\\
&\textcolor{blue}{+\frac{nQ\gamma^{m+n}\psi^3_\gamma\partial\psi^-_\gamma}{k\partial\gamma}+\frac{nQ\gamma^{m+n}\psi^-_\gamma\partial\psi^3_\gamma}{k\partial\gamma}+\frac{(m+1)nQ\gamma^{m+n-1}\psi^-_\gamma\psi^3_\gamma}{k}+\frac{(m+1)nQ\gamma^{m+n-1}\psi^-_\gamma\partial\psi^-_\gamma}{2k\partial\gamma}}\\
&+\frac{(m+1)m\gamma^{m+n-1}\partial\gamma}{Q}+mnQ\left[ \textcolor{red}{-\gamma^{m+n-1}\psi^-_\gamma\psi^3_\gamma-\gamma^{m+n-1}\partial\gamma}\textcolor{blue}{-\frac{\gamma^{m+n-1}\psi^-_\gamma\partial\psi^-_\gamma}{2k\partial\gamma}}\right.\\
&\left.\textcolor{blue}{+\frac{(m\gamma^{m-1}\partial\gamma\psi^-_\gamma+\gamma^m\partial\psi^-_\gamma)\gamma^{n-1}\psi^-_\gamma}{2k\partial\gamma}+\frac{\gamma^{m+n-1}\partial\gamma}{k}} \right]\\
&\textcolor{red}{+nQ\left[ -\gamma^{m+n-1}\psi^-\psi^3_\gamma+\gamma^{m+n+1}\psi^+\psi^3_\gamma+\gamma^{m+n}\psi^+\psi^-_\gamma+m\gamma^{m+n-1}\partial\gamma \right]}\\
&\textcolor{blue}{+\frac{nQ}{2k\partial\gamma}\left[ -\gamma^{m+n-1}\psi^-\partial\psi^-_\gamma+\gamma^{m+n+1}\psi^+\partial\psi^-_\gamma+(m\gamma^{m-1}\partial\gamma\psi^-+\gamma^m\partial\psi^-)\gamma^{n-1}\psi^-_\gamma\right.}\\
&\textcolor{blue}{\left.-2\gamma^{m+n}\psi^+\psi^-_\gamma\partial\gamma-((m-2)\gamma^{m-1}\partial\gamma\psi^++\gamma^{m-2}\partial(\gamma^2\psi^+))\gamma^{n+1}\psi^-_\gamma-2m\gamma^{m+n-1}(\partial\gamma)^2 \right]}\,.
\end{aligned}
\end{equation}
The terms in the same colour combine to give 
\begin{equation}
\begin{gathered}
\textcolor{red}{-n(m+n)Q\gamma^{m+n-1}\psi^-_\gamma\psi^3_\gamma}\,,\\
\textcolor{blue}{\frac{-n(m+n)Q\gamma^{m+n-1}\psi^-_\gamma\partial\psi^-_\gamma}{2k\partial\gamma}}\,.
\end{gathered}
\end{equation}
Thus, we see that
\begin{equation}
\begin{aligned}
&[\mathcal{L}^{susy}_m,\varphi^{susy}_n]\\
&=-n\oint_0dw\left( \gamma^{m+n}\partial\Phi+\frac{Q}{2}\frac{\gamma^{m+n}\partial^2\gamma}{\partial\gamma}+(m+n)Q\gamma^{m+n-1}\psi^-_\gamma\psi^3_\gamma+\frac{(m+n)Q\gamma^{m+n-1}\psi^-_\gamma\partial\psi^-_\gamma}{2k\partial\gamma} \right)\\
&\hspace{1cm}-\frac{m(m+1)\widetilde{Q}}{2}\delta_{m+n,0}\mathcal{I}\\
&=-n\varphi^{susy}_{m+n}-\frac{m(m+1)\widetilde{Q}}{2}\delta_{m+n,0}\mathcal{I}
\end{aligned}
\end{equation}
as expected.

\newpage

\bibliography{bibliography}

\providecommand{\href}[2]{#2}\begingroup\raggedright\begin{thebibliography}{10}

\bibitem{Eberhardt:2018ouy}
L.~Eberhardt, M.~R. Gaberdiel, and R.~Gopakumar, ``{The Worldsheet Dual of the
  Symmetric Product CFT},''
  \href{http://dx.doi.org/10.1007/JHEP04(2019)103}{{\em JHEP} {\bfseries 04}
  (2019) 103}, \href{http://arxiv.org/abs/1812.01007}{{\ttfamily
  arXiv:1812.01007 [hep-th]}}.

\bibitem{Eberhardt:2019ywk}
L.~Eberhardt, M.~R. Gaberdiel, and R.~Gopakumar, ``{Deriving the
  AdS$_{3}$/CFT$_{2}$ correspondence},''
  \href{http://dx.doi.org/10.1007/JHEP02(2020)136}{{\em JHEP} {\bfseries 02}
  (2020) 136}, \href{http://arxiv.org/abs/1911.00378}{{\ttfamily
  arXiv:1911.00378 [hep-th]}}.

\bibitem{Dei:2020zui}
A.~Dei, M.~R. Gaberdiel, R.~Gopakumar, and B.~Knighton, ``{Free field
  world-sheet correlators for ${\rm AdS}_3$},''
  \href{http://dx.doi.org/10.1007/JHEP02(2021)081}{{\em JHEP} {\bfseries 02}
  (2021) 081}, \href{http://arxiv.org/abs/2009.11306}{{\ttfamily
  arXiv:2009.11306 [hep-th]}}.

\bibitem{Fiset:2022erp}
M.-A. Fiset, M.~R. Gaberdiel, K.~Naderi, and V.~Sriprachyakul, ``{Perturbing
  the symmetric orbifold from the worldsheet},''
  \href{http://dx.doi.org/10.1007/JHEP07(2023)093}{{\em JHEP} {\bfseries 07}
  (2023) 093}, \href{http://arxiv.org/abs/2212.12342}{{\ttfamily
  arXiv:2212.12342 [hep-th]}}.

\bibitem{Gaberdiel:2021kkp}
M.~R. Gaberdiel, B.~Knighton, and J.~Vo\v{s}mera, ``{D-branes in AdS$_{3}$
  \texttimes{} S$^{3}$ \texttimes{} \ensuremath{\mathbb{T}}$^{4}$ at k = 1 and
  their holographic duals},''
  \href{http://dx.doi.org/10.1007/JHEP12(2021)149}{{\em JHEP} {\bfseries 12}
  (2021) 149}, \href{http://arxiv.org/abs/2110.05509}{{\ttfamily
  arXiv:2110.05509 [hep-th]}}.

\bibitem{Knighton:2024noc}
B.~Knighton, V.~Sriprachyakul, and J.~Vo\v{s}mera, ``{Topological defects and
  tensionless holography},'' \href{http://arxiv.org/abs/2406.03467}{{\ttfamily
  arXiv:2406.03467 [hep-th]}}.

\bibitem{Dei:2023ivl}
A.~Dei, B.~Knighton, and K.~Naderi, ``{Solving AdS$_{\boldsymbol 3}$ string
  theory at minimal tension: tree-level correlators},''
  \href{http://arxiv.org/abs/2312.04622}{{\ttfamily arXiv:2312.04622
  [hep-th]}}.

\bibitem{Eberhardt:2020bgq}
L.~Eberhardt, ``{Partition functions of the tensionless string},''
  \href{http://dx.doi.org/10.1007/JHEP03(2021)176}{{\em JHEP} {\bfseries 03}
  (2021) 176}, \href{http://arxiv.org/abs/2008.07533}{{\ttfamily
  arXiv:2008.07533 [hep-th]}}.

\bibitem{Gaberdiel:2023lco}
M.~R. Gaberdiel, R.~Gopakumar, and B.~Nairz, ``{Beyond the tensionless limit:
  integrability in the symmetric orbifold},''
  \href{http://dx.doi.org/10.1007/JHEP06(2024)030}{{\em JHEP} {\bfseries 06}
  (2024) 030}, \href{http://arxiv.org/abs/2312.13288}{{\ttfamily
  arXiv:2312.13288 [hep-th]}}.

\bibitem{Giveon:1998ns}
A.~Giveon, D.~Kutasov, and N.~Seiberg, ``{Comments on string theory on
  AdS(3)},'' \href{http://dx.doi.org/10.4310/ATMP.1998.v2.n4.a3}{{\em Adv.
  Theor. Math. Phys.} {\bfseries 2} (1998) 733--782},
  \href{http://arxiv.org/abs/hep-th/9806194}{{\ttfamily arXiv:hep-th/9806194}}.

\bibitem{Knighton:2024qxd}
B.~Knighton and V.~Sriprachyakul, ``{Unravelling AdS$_3$/CFT$_2$ near the
  boundary},'' \href{http://arxiv.org/abs/2404.07296}{{\ttfamily
  arXiv:2404.07296 [hep-th]}}.

\bibitem{Sriprachyakul:2024gyl}
V.~Sriprachyakul, ``{Superstrings near the conformal boundary of $\rm
  AdS_3$},'' \href{http://arxiv.org/abs/2405.03678}{{\ttfamily arXiv:2405.03678
  [hep-th]}}.

\bibitem{Eberhardt:2021vsx}
L.~Eberhardt, ``{A perturbative CFT dual for pure NS\textendash{}NS AdS$_{3}$
  strings},'' \href{http://dx.doi.org/10.1088/1751-8121/ac47b2}{{\em J. Phys.
  A} {\bfseries 55} no.~6, (2022) 064001},
  \href{http://arxiv.org/abs/2110.07535}{{\ttfamily arXiv:2110.07535
  [hep-th]}}.

\bibitem{Knighton:2023mhq}
B.~Knighton, S.~Seet, and V.~Sriprachyakul, ``{Spectral flow and localisation
  in AdS$_{3}$ string theory},''
  \href{http://dx.doi.org/10.1007/JHEP05(2024)113}{{\em JHEP} {\bfseries 05}
  (2024) 113}, \href{http://arxiv.org/abs/2312.08429}{{\ttfamily
  arXiv:2312.08429 [hep-th]}}.

\bibitem{Hikida:2023jyc}
Y.~Hikida and V.~Schomerus, ``{Engineering perturbative string duals for
  symmetric product orbifold CFTs},''
  \href{http://dx.doi.org/10.1007/JHEP06(2024)071}{{\em JHEP} {\bfseries 06}
  (2024) 071}, \href{http://arxiv.org/abs/2312.05317}{{\ttfamily
  arXiv:2312.05317 [hep-th]}}.

\bibitem{Balthazar:2021xeh}
B.~Balthazar, A.~Giveon, D.~Kutasov, and E.~J. Martinec, ``{Asymptotically free
  AdS$_{3}$/CFT$_{2}$},'' \href{http://dx.doi.org/10.1007/JHEP01(2022)008}{{\em
  JHEP} {\bfseries 01} (2022) 008},
  \href{http://arxiv.org/abs/2109.00065}{{\ttfamily arXiv:2109.00065
  [hep-th]}}.

\bibitem{Martinec:2021vpk}
E.~J. Martinec, ``{AdS3's with and without BTZ's},''
  \href{http://arxiv.org/abs/2109.11716}{{\ttfamily arXiv:2109.11716
  [hep-th]}}.

\bibitem{Kutasov:1999xu}
D.~Kutasov and N.~Seiberg, ``{More comments on string theory on AdS(3)},''
  \href{http://dx.doi.org/10.1088/1126-6708/1999/04/008}{{\em JHEP} {\bfseries
  04} (1999) 008}, \href{http://arxiv.org/abs/hep-th/9903219}{{\ttfamily
  arXiv:hep-th/9903219}}.

\bibitem{Dei:2021xgh}
A.~Dei and L.~Eberhardt, ``{String correlators on AdS$_{3}$: three-point
  functions},'' \href{http://dx.doi.org/10.1007/JHEP08(2021)025}{{\em JHEP}
  {\bfseries 08} (2021) 025}, \href{http://arxiv.org/abs/2105.12130}{{\ttfamily
  arXiv:2105.12130 [hep-th]}}.

\bibitem{Dei:2021yom}
A.~Dei and L.~Eberhardt, ``{String correlators on AdS$_{3}$: four-point
  functions},'' \href{http://dx.doi.org/10.1007/JHEP09(2021)209}{{\em JHEP}
  {\bfseries 09} (2021) 209}, \href{http://arxiv.org/abs/2107.01481}{{\ttfamily
  arXiv:2107.01481 [hep-th]}}.

\bibitem{Dei:2022pkr}
A.~Dei and L.~Eberhardt, ``{String correlators on $\text{AdS}_3$: Analytic
  structure and dual CFT},''
  \href{http://dx.doi.org/10.21468/SciPostPhys.13.3.053}{{\em SciPost Phys.}
  {\bfseries 13} no.~3, (2022) 053},
  \href{http://arxiv.org/abs/2203.13264}{{\ttfamily arXiv:2203.13264
  [hep-th]}}.

\bibitem{Polchinski:1998rq}
J.~Polchinski, \href{http://dx.doi.org/10.1017/CBO9780511816079}{{\em {String
  theory. Vol. 1: An introduction to the bosonic string}}}.
\newblock Cambridge Monographs on Mathematical Physics. Cambridge University
  Press, 12, 2007.

\bibitem{Gaberdiel:2022als}
M.~R. Gaberdiel, K.~Naderi, and V.~Sriprachyakul, ``{The free field realisation
  of the BVW string},'' \href{http://dx.doi.org/10.1007/JHEP08(2022)274}{{\em
  JHEP} {\bfseries 08} (2022) 274},
  \href{http://arxiv.org/abs/2202.11392}{{\ttfamily arXiv:2202.11392
  [hep-th]}}.

\bibitem{Ferreira:2017pgt}
K.~Ferreira, M.~R. Gaberdiel, and J.~I. Jottar, ``{Higher spins on AdS$_{3}$
  from the worldsheet},'' \href{http://dx.doi.org/10.1007/JHEP07(2017)131}{{\em
  JHEP} {\bfseries 07} (2017) 131},
  \href{http://arxiv.org/abs/1704.08667}{{\ttfamily arXiv:1704.08667
  [hep-th]}}.

\bibitem{Berenstein:1999gj}
D.~Berenstein and R.~G. Leigh, ``{Space-time supersymmetry in AdS(3)
  backgrounds},'' \href{http://dx.doi.org/10.1016/S0370-2693(99)00623-1}{{\em
  Phys. Lett. B} {\bfseries 458} (1999) 297--303},
  \href{http://arxiv.org/abs/hep-th/9904040}{{\ttfamily arXiv:hep-th/9904040}}.

\bibitem{Giveon:1999jg}
A.~Giveon and M.~Rocek, ``{Supersymmetric string vacua on AdS(3) x N},''
  \href{http://dx.doi.org/10.1088/1126-6708/1999/04/019}{{\em JHEP} {\bfseries
  04} (1999) 019}, \href{http://arxiv.org/abs/hep-th/9904024}{{\ttfamily
  arXiv:hep-th/9904024}}.

\bibitem{Eberhardt:2019qcl}
L.~Eberhardt and M.~R. Gaberdiel, ``{String theory on AdS$_3$ and the symmetric
  orbifold of Liouville theory},''
  \href{http://dx.doi.org/10.1016/j.nuclphysb.2019.114774}{{\em Nucl. Phys. B}
  {\bfseries 948} (2019) 114774},
  \href{http://arxiv.org/abs/1903.00421}{{\ttfamily arXiv:1903.00421
  [hep-th]}}.

\end{thebibliography}\endgroup
\bibliographystyle{utphys.sty}

\end{document}